\documentstyle{article}

\input amssym.def
 \input amssym.tex

\def\newpic#1{%
   \def\emline##1##2##3##4##5##6{%
      \put(##1,##2){\special{em:point #1##3}}%
      \put(##4,##5){\special{em:point #1##6}}%
      \special{em:line #1##3,#1##6}}}
\newpic{}
\def\R{\Bbb R}

\def\Z{\Bbb Z}

\def\C{\Bbb C}

\newtheorem{theorem}{Theorem}
\newtheorem{corollary}{Corollary}
\newtheorem{lemma}{Lemma}

\title{VERLINDE ALGEBRAS AND THE INTERSECTION FORM ON VANISHING CYCLES}
\author{S.M. Gusein-Zade \thanks{Supported in part by grants INTAS-94-4373
and RFBR-95-01-01122a. Independent University of Moscow, Russia.
E-mail: sabir\symbol{'100}ium.ips.ras.ru.} \and A. Varchenko
\thanks {Supported in part by NSF grant 9501290.
Address: Department of Mathematics, University of North Carolina,
Chapel Hill, N.C. 27599, USA. E-mail: av\symbol{'100}math.unc.edu.}}
\begin{document}


\def\Aut{{\operatorname{Aut}}}
\def\eps{\varepsilon}

\maketitle

\rightline{\large{\it Dedicated to E.Brieskorn on his sixtieth birthday}}

\begin{abstract}
We prove Zuber's conjecture {\cite{Z}} establishing connections
of the fusion rules of the $su(N)_k$ WZW model of conformal
field theory and the intersection form on vanishing cycles of
the associated fusion potential.
\end{abstract}

\section{Introduction}
We prove Zuber's conjecture {\cite{Z}} establishing connections
between the fusion rules of the $su(N)_k$ WZW model of conformal
field theory and the intersection form on vanishing cycles of
the associated fusion potential.

The fusion rules of the $su(N)_k$ model is the multiplication law
of the Verlinde algebra associated to the model.
The fusion rules can be encoded into a weighted graph, the Dynkin
diagram, see Section 2.6.
 The Dynkin diagram defines in a standard way a lattice
with a symmetric bilinear form, a group generated by reflections,
a Coxeter element of the group.

According to the Gepner theorem {\cite{Gep}} the Verlinde algebra can be
presented as a quotient algebra
$\C[x_1,...,x_{N-1}] / (\partial_iV_{N,k})$,
where $(\partial_iV_{N,k})$ is the ideal generated by the
first partial derivatives
of the fusion potential
$$
V_{N,k}(x)={(-1)^{N+k}\over (N+k)! \!}({d\over dt})^{N+k}
\log(1-tx_1+t^2x_2-\ldots+(-t)^{N-1}x_{N-1}+(-t)^N)\vert_{ t=0}.
$$
The fusion potential is a lower order deformation of a quasi-homogeneous
polynomial
$$
V^0_{N,k}(x)={(-1)^{N+k}\over (N+k)! \!}\left({d\over dt}\right)^{N+k}
\log(1-tx_1+t^2x_2-\ldots+(-t)^{N-1}x_{N-1})\vert_{ t=0}
$$
called the short fusion potential.
The short fusion potential has an isolated critical point at the origin.

We show that the lattice of the Verlinde algebra of $su(N)_k$
with the bilinear form, the reflection group, and the Coxeter
element is isomorphic to the lattice of vanishing cycles of the
critical point of the short fusion
potential together with the symmetric intersection
form, the monodromy group, and the operator of classical
monodromy, Theorem 3.

As a corollary we prove another Zuber's
conjecture in {\cite{Z}}, the level-rank
duality. We show that the lattice of $su(N+1)_k$
with the bilinear form, the reflection group, and the Coxeter
element is isomorphic to the lattice of $su(k+1)_N$
with the bilinear form, the reflection group, and the Coxeter
element. This conjecture was motivated in {\cite{Z}} by analogies
with the $N=2$ superconformal theories.

The second author thanks V.Petkova for attracting his attention to
articles {\cite{Gep}} and {\cite{Z}} and J.-B. Zuber for explaining
his conjectures and related results. This work would not be
written without his help.
The authors thank the IHES and the
organizers of the Singularity conference at Oberwolfach, July 1996,
for hospitality and stimulating atmosphere. The authors thank
R.-O.Buchweitz and W.Ebeling for useful discussions.

\section{Verlinde algebras}

The Verlinde algebra or the fusion ring is a finite dimensional
commutative associative algebra assigned to a model of
conformal field theory. It is an analog of the ring of
finite dimensional representations of a simple Lie algebra.
In this section we define the Verlinde algebra of the
$su(N)_k$ WZW model of conformal field theory.
We define its fusion potential, bilinear form, Dynkin
diagram and reflection group. We follow here {\cite{Gep}} and {\cite{Z}}.

\subsection{The representation ring of $su(2)$}

Finite dimensional representations of the Lie algebra $su(2)$
are labelled by non-negative integers, highest weights.
The representation with highest weight $m \in \Z_{\geq 0}$
has dimension $m+1$. Denote it by $[m]$. We have
\begin{equation}\label{eq1}
[n]\otimes[m]=
[m+n]\oplus[m+n-2]\oplus[m+n-4]\oplus\ldots\oplus[\vert m-n\vert].
\end{equation}
In particular, $[1]\otimes[m]=[m+1]\oplus[m-1]$ if $m>0$.

\subsection{The Verlinde algebra of the $su(2)_k$ model}

The Verlinde algebra $R=R(su(2)_k)$
of $su(2)$ at level $k$
is generated by the elements $[0], [1], \cdots, [k]$
as a vector space.
The multiplication (the fusion rules) is defined by the formula
$$
[n]\times[m]=
\sum_{{\ell=\vert m-n\vert\atop {m+n-\ell\mbox{\small{\
even}}}}}^{\min(2k-m-n, m+n)}
[\ell].
$$

This product is equal to the product defined by (1) if $m+n\leq k$
and does not contain representations with top highest weights
otherwise. In particular, we have
\begin{equation}\label{eq2}
\begin{array}{ll}
\mbox{$[1]\times[0]=[1]$,} & {}
\\
\mbox{$[1]\times[m]=[m-1]+[m+1]$,} & {0<m<k,}
\\
\mbox{$[1]\times[k]=[k-1]$.} & {}
\end{array}
\end{equation}
These formulae show that the Verlinde algebra is generated by
the elements $[0]$ and $ [1]$.

It is convenient to represent the formulae (\ref{eq2})
by a fusion graph, Figure 1.
The edges of the graph connecting the vertex $[m]$ with vertices
$[m-1]$ and $[m+1]$ represent the fact that
$[m-1]$ and $[m+1]$ enter $ [1]\times [m]$ with coefficient 1.
The element $[0]$ is the unit element of the Verlinde algebra.

\begin{figure}
$$
\unitlength=1.00mm
\special{em:linewidth 0.4pt}
\linethickness{0.4pt}
\begin{picture}(94.00,12.00)
\emline{10.00}{10.00}{1}{30.00}{10.00}{2}
\emline{32.00}{10.00}{3}{34.00}{10.00}{4}
\emline{34.00}{10.00}{5}{34.00}{10.00}{6}
\emline{36.00}{10.00}{7}{38.00}{10.00}{8}
\emline{40.00}{10.00}{9}{60.00}{10.00}{10}
\put(10.00,10.00){\circle*{1.00}}
\put(19.00,10.00){\circle*{1.00}}
\put(28.00,10.00){\circle*{1.00}}
\put(42.00,10.00){\circle*{1.00}}
\put(51.00,10.00){\circle*{1.00}}
\put(60.00,10.00){\circle*{1.00}}
\emline{60.00}{10.00}{11}{62.00}{10.00}{12}
\emline{64.00}{10.00}{13}{66.00}{10.00}{14}
\emline{68.00}{10.00}{15}{70.00}{10.00}{16}
\emline{72.00}{10.00}{17}{83.00}{10.00}{18}
\put(74.00,10.00){\circle*{1.00}}
\put(83.00,10.00){\circle*{1.00}}
\put(10.00,5.00){\makebox(0,0)[cc]{\footnotesize [0]}}
\put(19.00,5.00){\makebox(0,0)[cc]{\footnotesize [1]}}
\put(28.00,5.00){\makebox(0,0)[cc]{\footnotesize [2]}}
\put(42.00,5.00){\makebox(0,0)[cc]{\footnotesize [m-1]}}
\put(51.00,5.00){\makebox(0,0)[cc]{\footnotesize [m]}}
\put(60.00,5.00){\makebox(0,0)[cc]{\footnotesize [m+1]}}
\put(74.00,5.00){\makebox(0,0)[cc]{\footnotesize [k-1]}}
\put(83.00,5.00){\makebox(0,0)[cc]{\footnotesize [k]}}
\end{picture}
$$
\caption{Fusion graph of $su(2)_k$, Dynkin diagram $A_{k+1}$.} \label{fig1}
\end{figure}
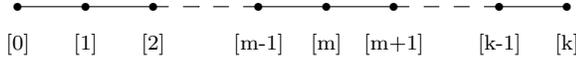

Let us express an element $[m]$ as a polynomial of the
element
 $[1]$. Denote the element
$[1]$ by $x$ and the element $[0]$ by 1. Using (\ref{eq2}) we get
$[2]=x^2-1, \, [3]=x^3-2x.$ More generally, if
$[m]= P_m(x)$, then
$$
xP_m(x)=P_{m-1}(x)+P_{m+1}(x).
$$
The solution to this recursive relation
with the initial condition $P_0(x)=1$ and $P_1(x)=x$
 is given by
the Chebyshev polynomials of the second kind,
$$
P_m(2\mbox{cos}\theta)={\mbox{sin}(m+1)\theta \over
\mbox{sin}\theta }.
$$
According to Gepner {\cite{Gep}}, the Verlinde algebra of $su(2)_k$
is the quotient algebra
$$
R(su(2)_k) = {\C[x] \over (dV_{2,k}/dx)}
$$
where $(dV_{2,k}/dx)$ is the ideal generated by the derivative
of the fusion potential $V_{2,k}(x)$, and the fusion potential
$V_{2,k}(x)$ is the Chebyshev polynomial of the first kind,
$$
(2+k)\,V_{2,k}(2\mbox{cos}\theta)\,=\,2\mbox{cos}(2+k)\theta.
$$

\smallskip
{\bf Remark.} The Chebyshev polynomial of the first kind is
a polynomial with non-degenerate critical
points and only two critical values.

\smallskip
{\bf Remark.} The fusion graph of the multiplication by element
$[1]$ is the Dynkin
diagram of type $A_{k+1}$. At the same time the fusion
potential $V_{2,k}$ is a lower order deformation of the monomial
$x^{k+2}$ which has a critical point of type $A_{k+1}$
in the terminology of singularity theory.

\subsection{The representation ring of $su(N)$}

Let $\Lambda_1$, $\Lambda_2$, $\dots,$ $\Lambda_{N-1}$ be the fundamental
weights of the Lie algebra $su(N)$. They are linear independent vectors in
the dual to the Cartan subalgebra of $su(N)$,
$$
\frak h^*=\R\{e_1, e_2, \dots, e_N\}/\R\{e_1+e_2+\cdots+e_N\}.
$$
 The vector $\Lambda_i$ is the image of the vector
$(1_1,...,1_i, 0, \dots, 0).$
An irreducible finite dimensional representation
of $su(N)$ is determined by its highest weight
$\lambda=\lambda_1\Lambda_1+\cdots+\lambda_{N-1}\Lambda_{N-1}$, where
$\lambda_i$ are non-negative integers. Let $[\lambda]$ denote
the irreducible
representation with highest weight $\lambda$.
 The tensor product $[\lambda]\otimes[\mu]$
of two representations is a sum of irreducible representations.
The representation ring of $su(N)$ is generated by representations
$[0]$, $[\Lambda_1]$, ..., $[\Lambda_{N-1}]$.

The Verlinde algebra of $su(N)_k$ is obtained from the
representation ring by suitable
cutting off representations with top highest weights.

\subsection{Verlinde algebra of the $su(N)_k$ model}

The Verlinde algebra $R=R(su(N)_k)$ of the Lie algebra
$su(N)$ at level $k$ is a finite dimensional commutative associative
algebra generated as a vector space by elements $[\lambda]$ of the set
$$
P_{N,k}=
\{\sum_{i=1}^{N-1}
\lambda_i\Lambda_i:\ \lambda_i\in\Z_{\ge0},\
\sum_{i=1}^{N-1} \lambda_i\le k\}.
$$
The multiplication is defined by the fusion rules.
We use the formula in {\cite{K}}, Sec. 13.35, as the definition of
the multiplication, see the same formula in
{\cite{Good}} and {\cite{W}}.
The element $[0]$ is the unit element of the algebra. As an algebra,
the Verlinde algebra is generated by the elements
$[0]$, $[\Lambda_1]$, ...,$[\Lambda_{N-1}]$. To define
the multiplication it is enough to define the multiplication
by the generators $[\Lambda_1]$, ...,$[\Lambda_{N-1}]$.

Let $e_1=\Lambda_1$,
$e_2=\Lambda_2-\Lambda_1$, \dots, $e_{N-1}=\Lambda_{N-1}-\Lambda_{N-2}$,
$e_N=-\Lambda_{N-1}$ . The vectors $e_1$, $e_2$, \dots,$e_N$ are linear
dependent: $e_1+e_2+\dots+e_N=0$. For any $p=1,...,N-1$, we have
$$
[\Lambda_p]\times[\lambda]=\sum \, [\mu]
$$
where the sum is over all $\mu \in P_{N,k}$ such that
$\mu=\lambda+e_{i_1}+\cdots+e_{i_p}$,
$1\leq i_1< ... <i_p \leq N$.

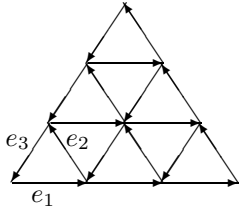
\begin{figure}
$$
\unitlength=1.00mm
\special{em:linewidth 0.4pt}
\linethickness{0.4pt}
\begin{picture}(40.00,36.00)
\put(10.00,10.00){\vector(1,0){10.00}}
\put(20.00,10.00){\vector(1,0){10.00}}
\put(30.00,10.00){\vector(1,0){10.00}}
\put(15.00,18.00){\vector(-2,-3){5.27}}
\put(25.00,18.00){\vector(-2,-3){5.27}}
\put(35.00,18.00){\vector(-2,-3){5.27}}
\put(20.00,10.00){\vector(-2,3){5.33}}
\put(30.00,10.00){\vector(-2,3){5.33}}
\put(40.00,10.00){\vector(-2,3){5.33}}
\put(15.07,18.00){\vector(1,0){10.00}}
\put(25.00,18.00){\vector(1,0){10.07}}
\put(20.00,26.00){\vector(-2,-3){5.33}}
\put(30.00,26.00){\vector(-2,-3){5.33}}
\put(25.00,18.00){\vector(-2,3){5.27}}
\put(35.00,18.00){\vector(-2,3){5.27}}
\put(20.00,26.00){\vector(1,0){10.00}}
\put(30.00,26.00){\vector(-2,3){5.33}}
\put(25.07,34.00){\vector(-2,-3){5.33}}
\put(14.00,8.00){\makebox(0,0)[cc]{$e_1$}}
\put(17.00,15.00){\makebox(0,0)[lb]{$e_2$}}
\put(9.00,15.00){\makebox(0,0)[lb]{$e_3$}}
\end{picture}
$$
\caption{Fusion graph $\gamma_1$ of $su(3)_k$, $k=3$.}\label{fig2}
\end{figure}

The multiplication by an element $[\Lambda_p]$ is useful to represent
by a fusion graph $\gamma_p$ and by an adjacency matrix $G_p$.
The fusion graph $\gamma_p$ is an oriented graph
with the vertex set $P_{N,k}$ and
such that an edge goes from a vertex $\lambda$
to a vertex $\mu$ if and only if
$\mu=\lambda+e_{i_1}+\cdots+e_{i_p}$,
$1\leq i_1< ... <i_p \leq N$.
The fusion graph $\gamma_1$ of $su(3)_k$ is shown in Figure
\ref{fig2}.

For any $p=1,...,N-1$ the fusion graph $\gamma_{N-p}$
is obtained from the fusion graph $\gamma_{p}$
by reversing all the orientations of the edges.

The set of vertices $P_{N,k}$ has a natural $\Z_N$-valued grading:
$$
\lambda \qquad \mapsto \qquad
\tau(\lambda)=\sum_{n=1}^{N-1}n\,\lambda_n\ \mbox{mod\,} N.
$$
If an edge of the fusion graph $\gamma_p$ goes from a vertex $\lambda$
to a vertex $\mu$, then $\tau(\mu)=\tau(\lambda)+p$.

The adjacency matrix $G_p$ is a matrix with entries
$(G_p)_{\lambda,\mu}$ where $\lambda, \mu \in P_{N,k}$.
The entries of the adjacency matrix are given by the formula:
$(G_p)_{\lambda,\mu}=1$ if
$\mu=\lambda+e_{i_1}+\cdots+e_{i_p}$,
$1\leq i_1< ... <i_p \leq N$, and
$(G_p)_{\lambda,\mu}=0$ otherwise.

If $(G_p)_{\lambda,\mu}=1$ , then $\tau(\mu)=\tau(\lambda)+p$.

For any $p=1,...,N-1$ the matrix $G_{N-p}$ is the transpose
of the matrix $G_p$, $G_{N-p}=(G_p)^t$.

\subsection{The fusion potential of $su(N)_k$}

Denote the elements $[\Lambda_1]$, ..., $[\Lambda_{N-1}]$
of the Verlinde algebra
by
$x_1$, ..., $x_{N-1}$, respectively, and the element $[0]$ by 1.
Then any element of the Verlinde algebra can be written as a polynomial
in $x_1$, ..., $x_{N-1}$, see in {\cite{Gep}} a general formula. These
polynomials
are multidimensional analogs of the Chebyshev polynomials of the
second kind. For instance, for $su(3)_k$ we have
$[2\Lambda_1]=(x_1)^2-x_2$,
$[\Lambda_1 + \Lambda_2]=x_1x_2-1$,
$[2\Lambda_2]=(x_2)^2-x_1$, ...

There is a beautiful Gepner's theorem.

\begin{theorem}\label{theo1} {\rm ({\cite{Gep}})}
The Verlinde algebra of $su(N)_k$ is the quotient algebra
$$
R(su(N)_k) = {\C[x] \over (\partial_iV_{N,k})}
$$
where $(\partial_iV_{N,k})$ is the ideal generated by the
first partial derivatives
of the fusion potential $V_{N,k}(x)$, and the fusion potential
is a polynomial in $x_1$, ..., $x_{N-1}$ given below.
\end{theorem}

There are three formulae for the fusion potential
$V_{N,k}(x)$. Let $h=N+k$. According to the first formula,
$$
V_{N,k}(x)={(-1)^{h}\over h! \!}\left({d\over dt}\right)^{h}
\log(1-tx_1+t^2x_2-\ldots+(-t)^{N-1}x_{N-1}+(-t)^N)\vert_{ t=0}.
$$
According to the second, the fusion potential $V_{N,k}(x)$ is
the symmetric polynomial $ \sum_{i=1}^N (y_i)^{h}/h $ written
as a polynomial in the
elementary symmetric functions $x_1$, ..., $x_N$,
\begin{equation}\label{eq4}
x_n=\sum_{1\leq i_1<...<i_n\leq N} y_{i_1}\dots y_{i_n},
\end{equation}
where we assume that $x_N=1$. Finally, the multiple
$(N+k)\,V_{N,k}(x)$ of the fusion potential can be written
as the determinant of an $h\times h$ matrix $D$ with
non-zero elements given by
$$
\begin{array}{ll}
D_{n,n+1} = 1 & \quad n=1,...,h-1,
\\
D_{n,1} = n \,x_n & \quad n=1,...,h,
\\
D_{i+n-1,i} = x_n & \quad n=1,...,h-1,\,
i=2,...,h-n+1,
\end{array}
$$
where we assume that $x_N=1$ and $x_n=0$ for $n>N$.
(One can find this formula for the polynomial
$ \sum_{i=1}^N (y_i)^h $ in {\cite{M}}.)

An important role is played by the top quasi-homogeneous part
of the fusion potential. Namely, consider
the polynomial
\begin{equation}\label{eq6}
V^0_{N,k}(x)={(-1)^h\over h! \!}\left({d\over dt}\right)^h
\log(1-tx_1+t^2x_2-\ldots+(-t)^{N-1}x_{N-1})\vert_{ t=0}.
\end{equation}

The polynomial $V^0_{N,k}(x)$ can be defined as
the symmetric polynomial\linebreak
$ \sum_{i=1}^{N-1} (y_i)^h/h$ written
as a function in $x_1$, ..., $x_{N-1}$.
Finally, $(N+k)\,V^0_{N,k}(x)$ is equal to the determinant of the
$h\times h$ matrix $D$ with non-zero elements given
by
$$
\begin{array}{ll}
D_{n,n+1}=1 & \quad n=1,...,h-1,
\\
D_{n,1}=n \,x_n & \quad n=1,...,h,
\\
D_{i+n-1,i}= x_n & \quad n=1,...,h-1,\,
i=2,...,h-n+1,
\end{array}
$$
where we assume that $x_n=0$ for $n>N-1 $.

The polynomial $V_{N,k}^0$ will be called the short fusion potential.
The short fusion potential is a quasi-homogeneous polynomial
of degree $h$ with the weight of the variable $x_n$ equal to $n$,
$n=1,..., N-1$.
The fusion potential $V_{N,k}$ is a lower order deformation of the short
fusion potential by terms of degree less than $h$.

The short fusion potential has an isolated critical point
at the origin, since the polynomial
$ \sum_{i=1}^{N-1} (y_i)^h $
has an isolated critical point at the origin.
The Milnor number of the critical point of the short fusion potential
is equal to the dimension of the Verlinde algebra.

\smallskip
{\bf Example.} For $su(3)_1$ , $4V_{3,1}(x_1,x_2)$ $=$
$(x_1)^4-4(x_1)^2x_2+4x_1+2(x_2)^2$, $4V^0_{3,1}(x_1,x_2) =
(x_1)^4-4(x_1)^2x_2+2(x_2)^2$. The critical point of
the short fusion potential has type $A_3$,
see in {\cite{AGV}} the classification of critical points.

\smallskip
{\bf Example.} For $su(3)_2$,
$5V_{3,2}(x_1,x_2)=(x_1)^5-5(x_1)^3x_2+5x_1(x_2)^2
+5(x_1)^2-5x_2$, $5V^0_{3,2}(x_1,x_2)
=(x_1)^5-5(x_1)^3x_2+5x_1(x_2)^2$.
The critical point of the short fusion potential has type $D_6$.

\smallskip
{\bf Remark.} The fusion potential $V_{N,k}(x_1, ..., x_{N-1})$
is a multidimensional analog of the Chebyshev polynomials of
the first kind. The fusion potential has the following
remarkable property (which we will not use).
Namely, the fusion potential has only $N$ critical values,
and the critical values are $N-$th
roots of unity multiplied by $N/h$.

In fact, a critical point of the fusion potential corresponds
to a critical point of the function
$ \sum_{i=1}^N (y_i)^{h}/h $ restricted to the hypersurface
 $y_1\cdots y_N=1$. Such a critical point corresponds
to a critical point of the function
$ \lambda (1-y_1\cdots y_N)/h + \sum_{i=1}^N (y_i)^{h}/h $ ,
where $\lambda$ is the Lagrange multiplier. The critical points
of the last function are solutions to the system of equations
$y_1\cdots y_N=1$ and $y_n^h=\lambda/h$, $n=1, ..., N$.
The system implies that $\lambda^N=h^N$ and the critical value
is equal to $N\lambda/h^2$.

 Notice also that there is another lower order deformation
of the short fusion potential with only $N$ critical
values. That polynomial is the direct sum
of the Chebyshev polynomials of one variable,
$\sum_{i=1}^{N-1} V_{2,h-2}(y_i)$, written as a function
of the elementary symmetric functions $x_n$.
Its critical values are $-2(N-1)/h$, $-2(N-3)/h$, \dots,
$2(N-3)/h$, $2(N-1)/h$.

\smallskip
{\bf Remark.}
There is a general problem of finding the maximal number of
critical points with the same critical value of a lower order deformation
of a quasihomogeneous polynomial of given degree and weights,
see {\cite{AGV}}, Sec. 14.3.2, {\cite{Ch}}, {\cite{Gor}} and references therein.

The direct sum
$\widetilde V =\sum_{i=1}^{N-1} V_{2,h-2}(y_i)$
of the Chebyshev polynomials of the first kind,
written as a function of the elementary symmetric functions $x_n$,
provides an example of a lower order deformation
of a quasihomogeneous polynomial of degree $h$ and weights
$1, 2, ... , N-1$ with many critical points on
one level.
Namely, denote by $\#$ the number of critical points of
this function with critical value $0$ for $N$
odd and with critical value
$-2/h$ for $N$ even.
Then $\#$ is equal to
$$
\begin{array}{cl}
\left({h/2}\atop{N/2}\right)\left({h/2-1}\atop{N/2-1}\right)
& \mbox{{}\qquad for $N$ and $h$ even;}
\\
\left({(h-1)/2}\atop{N/2}\right)\left({(h-1)/2}\atop{N/2-1}\right)
& \mbox{{}\qquad for $N$ even, $h$ odd;}
\\
\left({h/2}\atop{(N-1)/2}\right)\left({h/2-1}\atop{(N-1)/2}\right)
& \mbox{{}\qquad for $N$ odd, $h$ even;}
\\
\left({(h-1)/2}\atop{(N-1)/2}\right)\left({(h-1)/2}\atop{(N-1)/2}\right)
& \mbox{{}\qquad for $N$ and $h$ odd,}
\end{array}
$$
while the total number of critical points is equal to
$\left({h-1}\atop{N-1}\right)$.

Considering sums of Chebyshev polynomials of the first kind
is a standard way to construct polynomials with many critical points
on one level .
In particular, if $h$ is divisible by $2$, $3$, ..., $(N-1)$, then
a sum of the form
\begin{equation}\label{eqnew}
\sum_{i=1}^{N-1} \pm V_{2, h/i-2}(x_i) / i
\end{equation}
gives another example of a lower order deformation of a quasihomogeneous polynomial
of degree $h$ with weights of variables equal to
$1$, $2$, $3$, ..., $(N-1)$ and many critical points on one level.

It turns out that the maximal
number of critical points on one level of the
function $\widetilde V$
is greater than the maximal number of critical points on one level
of the sums of Chebyshev polynomials indicated in (\ref{eqnew}).
For example, for $N=3$, i.e. for functions of two variables, and $h=2k$ we
have
$\#=k^2-k$, while the maximal number of critical points of functions
(\ref{eqnew}) on one level
is $k^2-(3/2)k+1$ for $k$ even and
$k^2-(3/2)k+(1/2)$ for $k$ odd.
For $N=4$, i.e. for functions of three variables, and $h=6k$
we have $\#={1\over 2}(27k^3-18k^2+3k)$, while
the maximal number of critical
points of functions (\ref{eqnew}) on one level
is equal to ${1\over 2}(27k^3-18k^2)$ for $k$ even and
to ${1\over 2}(27k^3-21k^2+4k)$ for $k$ odd, and so on.

\subsection{\hspace{-2mm}The Dynkin diagram and the reflection group of $su(N)_k$}

Consider a lattice $L=L(su(N)_k)$ over $\Z$ with a basis $\alpha_\lambda$,
$\lambda \in P_{N,k}$. Introduce a bilinear symmetric form
$B_{N,k}$ on $L$,
\begin{equation}\label{eq8}
B_{N,k}\left(\alpha_\lambda,\alpha_\mu\right)=
2\delta_{\lambda\mu}+
(G_1+G_2+...+G_{N-1})_{\lambda\mu}.
\end{equation}
The bilinear form can be described by its Dynkin diagram.
The vertices of the Dynkin diagram are
elements of $P_{N,k}$. Any two
 vertices $\lambda$ and $\mu$ are connected
by an edge if and only if
$B_{N,k}\left(\alpha_\lambda,\alpha_\mu\right)\ne 0.$
If $B_{N,k}\left(\alpha_\lambda,\alpha_\mu\right)\ne 0$,
then the vertices are connected by an edge of multiplicity
$-B_{N,k}\left(\alpha_\lambda,\alpha_\mu\right)$.
This graph will be called the Dynkin diagram of
the $su(N)_k$ model.

According to the definition of the adjacency matrices,
all edges of the Dynkin diagram have multiplicity $-1$.
Two edges $\lambda$ and $\mu$ are connected by an edge
if and only if $\mu=\lambda+e_{i_1}+\cdots+e_{i_p}$,
$1\leq i_1< ... <i_p \leq N$, for some $p,\,
0< p <N .$

Fix a
lexicographical ordering of the set of vertices
of the Dynkin diagram. We assume that $\lambda < \mu$
if there is $i$ such that $\lambda_n = \mu_n$
for $n=i+1,...,N-1$ and $\lambda_i < \mu_i$.

The Dynkin diagram can be obtained from the fusion graphs
$\gamma_1,$ $ \dots,$ $\gamma_{N-1}$.
Namely, the Dynkin diagram and the
fusion graphs have the same set of vertices. The union of edges
of the graphs $\gamma_1$, ..., $\gamma_{N-1}$ is naturally decomposed
into pairs of edges with the opposite orientations. Forgetting the
orientation we construct from each pair
an edge of multiplicity $-1$ of the Dynkin diagram.

Notice also that knowing the Dynkin diagram we can reconstruct
the fusion graphs, since an edge of the fusion graph $\gamma_p$
goes from $\lambda$ to $\mu$ only if $\tau(\mu)=\tau(\lambda)+p$.

Thus, the Dynkin diagram encodes the operators
of multiplication
by $[\Lambda_1]$, ...,
$[\Lambda_{N-1}]$ in the Verlinde algebra.

Introduce the group $\Gamma_{N,k}$ as the group of linear automorphisms
of the lattice $L$ generated by reflections $s_\lambda$ at the
hyperplanes orthogonal to the
basis elements $\alpha_\lambda, \lambda \in P_{N,k}$,
$$
s_{\lambda}\, :\,x\,\mapsto \,x-(x,\alpha_\lambda)\alpha_\lambda.
$$
The group $\Gamma_{N,k}$ will be called the reflection group of $su(N)_k$.

The product of all basis reflections,
$$
M_{N,k} = s_{0} \cdots s_{k\Lambda_{N-1}} =
\prod_{\lambda \in P_{N,k}} s_{\lambda},
$$
written in the lexicographical order will be
 called the Coxeter element
of the reflection group of the Verlinde algebra.

Notice that this definition of the Coxeter element differs
from Zuber's definition in {\cite{Z}}.

\begin{theorem}\label{theo2} {\rm {\cite{Z}}}.
The form $B_{N,k}$ is positive definite if and only if
$N=2$ or $N=3,\,k=0, 1, 2,$ or $k=N, N+1$.
\end{theorem}

The finite reflection groups $\Gamma_{2,k}$, $\Gamma_{3,0}$,
$\Gamma_{3,1}$, $\Gamma_{3,2}$, $\Gamma_{N,N}$, $\Gamma_{N,N+1}$ have types
$A_{k+1}$, $A_{1}$,
$A_{3}$, $D_{6}$, $A_{1}$, $A_{N+1}$,
respectively. For other pairs $N, k$ the groups $\Gamma_{N,k}$ are
infinite.

It turns out that the bilinear form $B_{N,k}$, the reflection
group $\Gamma_{N,k}$ and the Coxeter element $M_{N,k}$
of the $su(N)_k$ model can be defined in terms of the
topology of the level hypersurfaces of the short fusion
potential of $su(N)_k$. We recall the necessary definitions in the next
section.

\section{\hspace{-1mm}The intersection form and the monodro\-my gro\-up
of an isolated critical point, [AGV]}

Let $f:(\C^n,0)\to(\C,0)$ be a germ of a holomorphic function
with an isolated critical point at the origin.
Let $B_\delta \subset \C^n$ denote the ball of radius $\delta$
with the center at the origin.
Fix small
$\delta\gg\vert\eps\vert>0$,
$\eps \in \C$.
The manifold
$V_\eps=f^{-1}(\eps)\cap B_\delta$ is called the Milnor fiber of
the germ $f$. The Milnor fiber is homotopy equivalent to the bouquet of
$(n-1)$-dimensional spheres. The number $\mu=\mu(f)$
of the spheres is called the Milnor number of $f$.
The middle homology group
$L_f=H_{n-1}(V_\eps; \Z)$ of the Milnor fiber is called the Milnor lattice.
The intersection form is a bilinear form on the Milnor lattice.
The intersection form is symmetric for $n$ odd and is skew-symmetric
for $n$ even.


Consider a deformation $f_t : \C^n \to \C$, $t\in [0,1]$,
of the germ $f$ such that for small non-zero $t$ the
functions $f_t$ have only non-degenerate critical points with
pair-wise distinct critical values. The number of critical points
equals the Milnor number.
Fix such $t_0$, and set
$\widetilde f=f_{t_0}(x)$. Let $p_1$, $p_2$, \dots, $p_\mu$
denote the critical points of $\widetilde f$, and let $z_i=\widetilde f(p_i)$
denote the critical values.

Let $\widetilde V_z=\widetilde f^{-1}(z)\cap B_\delta$ denote the
local level set of the function $\widetilde f$.
Let $z_0$ be a non--critical value of the function $\widetilde f$ such that
$\vert z_0\vert>\vert z_i\vert$ for $i=1, 2, \cdots, \mu$.
The manifold $\widetilde V_{z_0}$
is diffeomorphic to the Milnor fiber $V_\eps$ of the germ $f$.

Let $u_i$, $i=1, 2, \dots, \mu$, be smooth
non-self-intersecting paths
connecting the critical values $z_i$ with the non-critical value
$z_0$ and such that $u_i(0)=z_i$, $u_i(1)=z_0$. We assume that
the paths lie inside the circle
$\{z \in \C \, : \, \vert z \vert\le\vert z_0\vert\}$
and any two of them intersect only at the point $z_0$.
We enumerate the paths (and thus the critical
values and critical points) in the order they
enter the point $z_0$ counting clockwise and
starting from the boundary of the circle.

By the Morse lemma, for each critical point $p_i$ of the function
$\widetilde f$ there exists a system of local
coordinates $y_1$, $y_2$, \dots,
$y_n$ centered at $p_i$ such that the function $\widetilde f$
can be written in the form
$\widetilde f(y_1, y_2, \ldots, y_n)=z_i+\sum_{j=1}^{n} (y_j)^2$.
For a small $\tau$, the level manifold $\widetilde V_{u_i(\tau)}$
contains the $n-1$-dimensional sphere $ S_i(\tau)$
given by the equations $\sum_{j=1}^n (y_j)^2+z_i= u_i(\tau)$,
${\rm{Im}}\,\left(y_j/(u_i(\tau)-z_i)^{1/2}\right) \,=\,0$,
$j=1, ..., n$. For $\tau=0$ the sphere degenerates
to the critical point $p_i$. Lifting the homotopy of $\tau$ from
0 to 1, we construct a family of $(n-1)$-dimensional spheres
$S_i(\tau) \subset\widetilde V_{u_i(\tau)}$
for all $\tau$ between 0 and 1.
The homology class $\delta_i\in H_{n-1}(\widetilde V_{z_0};\Z)$
defined (up to orientation) by the sphere $S_i(1)$ is called the
vanishing cycle corresponding to the path $u_i$.
The vanishing cycles $\delta_1$, $\delta_2$, \dots, $\delta_\mu$ form
a basis of the homology group
$H_{n-1}(\widetilde V_{z_0};\Z)\cong H_{n-1}(V_\eps; \Z)=L_f$.
A basis constructed in this way is called distinguished.
The distinguished basis depends on the choice of the paths
$u_i$.

The system of non-singular level manifolds $\widetilde V_z$
forms a locally trivial bundle over the complement
to the set of critical values,
$\C\setminus\{z_1, z_2, \ldots, z_\mu\}$.
Any loop in the complement
with the end points at $z_0$ can be lifted to an isotopy of
the fibers over the loop. The isotopy induces a linear
automorphism of the homology group
$H_{n-1}(\widetilde V_{z_0})\cong L_f$ called the monodromy
transformation. The set of the monodromy transformations corresponding
to all loops forms a group called the monodromy
group of the germ $f$.

The set of paths $u_i$ , $i=1, 2, \dots, \mu$, gives
a set of generators, $s_i:L_f\to L_f$, of the monodromy group,
$$
s_i\, : a \, \mapsto \,a+(-1)^{n(n+1)/2}(a, \delta_i)\delta_i,
$$
here $(a, \delta_i)$ is the intersection number of the cycles
$a$ and $\delta_i$. The transformation $s_i$ is called
the Picard-Lefschetz transformation. Thus, the monodromy group
is determined by the intersection form and a distinguished basis.

We have $(\delta_i, \delta_i)=(-1)^{(n-1)(n-2)/2}(1+(-1)^{n-1})$.
This means that this self-intersection number
is equal to $0$ for even $n$ and to $(-1)^{(n-1)/2}$ for $n$ odd.
Hence, for an odd number of variables, $n$,
a Picard-Lefschetz transformation,
$s_i$, is the reflection at the hyperplane orthogonal
to the vanishing cycle $\delta_i$.

The monodromy transformation corresponding to the path going
counterclockwise around all the critical values is called
the operator of classical monodromy. It equals the product
$s_1\circ s_2\circ\cdots \circ s_\mu$
of the Picard-Lefschetz transformations.

Let $f:(\C^n, 0)\to(\C,0)$ and $g:(\C^m, 0)\to(\C,0)$ be
two germs of holomorphic functions of $n$ and $m$ variables,
respectively.
The germ of the function
$$
f\oplus g\, :\, (\C^{n+m},0) \, \to \, (\C,0),
\qquad (x,y)\,
\mapsto \, f(x)+g(y)\, ,
$$
is called the direct sum of the germs $f$ and $g$.
Let $\{\delta_i\}$, $i=1,\ldots,\mu(f)$, and
$\{\delta_j^\prime\}$, $j=1,\ldots,\mu(g)$, be
distinguished bases of vanishing cycles of the germs $f$ and $g$.
Gabrielov's theorem ({\cite{Gab}}, {\cite{AGV}}) describes a distinguished
basis of vanishing cycles
of the germ $f\oplus g$ and the corresponding intersection form.
Consider the lattice $L_f\otimes L_g$ with the basis
$\Delta_{ij}=\delta_i\otimes\delta_j^\prime$ ordered lexicographically:
$(i,j)<(i^\prime,j^\prime)$ if either $j<j^\prime$ or
$j=j^\prime$ and $i<i^\prime$. Introduce a bilinear form
on the lattice $L_f\otimes L_g$ by the formulae:
$$
\begin{array}{ll}
(\Delta_{i\,j_1},\Delta_{i\,j_2})=sgn(j_2-j_1)^n(-1)^{nm+n(n-1)/2}
(\delta_{j_1}^\prime,\delta_{j_2}^\prime)&{\mbox{$j_1\ne j_2$,}}
\\
(\Delta_{i_1j\,},\Delta_{i_2j\,})=sgn(i_2-i_1)^m(-1)^{nm+m(m-1)/2}
(\delta_{i_1},\delta_{i_2})&{\mbox{$i_1\ne i_2$,}}
\\
(\Delta_{i_1j_1},\Delta_{i_2j_2})
=0&{\mbox{$(i_2-i_1)(j_2-j_1)<0$,}}
\\
(\Delta_{i_1j_1},\Delta_{i_2j_2})=sgn(i_2-i_1)(-1)^{nm}
(\delta_{i_1},\delta_{i_2})(\delta_{j_1}^\prime,\delta_{j_2}^\prime)
&{\mbox{$(i_2-i_1)(j_2-j_1)>0$,}}
\\
(\Delta_{ij}, \Delta_{ij})=(-1)^{(n+m-1)(n+m-2)/2}(1+(-1)^{n+m-1}).
&{\mbox{}}
\end{array}
$$
Then there is a natural isomorphism of the lattice $L_f\otimes L_g$
and the Milnor lattice $L_{f\oplus g}$ of the direct sum sending the
bilinear form on the lattice $L_f\otimes L_g$ to the intersection
form and the basis $\Delta_{ij}$ to a distinguished basis.

Gabrielov's theorem also describes a system of paths
which defines the distinguished basis of $L_{f\oplus g}$
corresponding to $\{\Delta_{ij}\}$, see {\cite{Gab}}.

If $f=g:(\C^n,0)\to(\C,0)$, then $f\oplus f$ is a germ of a
function on $\C^n\times\C^n$. The permutation of the factors acts
on the Milnor lattice $L_{f\oplus f}$ by the formula
$\sigma_*(\delta_i\otimes\delta_j)=(-1)^n(\delta_j\otimes\delta_i)$.
This follows from the description in {\cite{AGV}}
of the Milnor fiber of a direct sum as the joint of the Milnor
fibers of summands.

Let $Q(y)$ be a non-degenerate quadratic form of variables
$y=(y_1,...,y_m)$. The germ of the function
$$
 f\oplus Q\, :\, (\C^{n+m},0) \, \to \, (\C,0),
\qquad (x,y)\,
\mapsto \, f(x)+Q(y)\, ,
$$
is called a stabilization of the germ $f$.
The Milnor number of the stabilization is equal to
the Milnor number of the initial germ. The corresponding
Milnor lattices are naturally isomorphic.
The intersection form and
the set of distinguished bases of the germ $f$ defines
the intersection form and the set of distinguished bases of
the stabilization.

Namely, there exists a natural isomorphism between the
Milnor lattices $L_f$ and $L_{f\oplus Q}$ which establishes
a one-to-one correspondence between
the distingui\-shed bases.
If $\{\delta_i\}$ is a distinguished basis of $L_f$
and $\{\widetilde\delta_i\}$ is the corresponding distinguished
basis of $L_{f\oplus Q}$ , then
\begin{equation}\label{eq9}
(\widetilde\delta_i,\widetilde\delta_j)=
[{\rm{sgn}}\,(j-i)]^m(-1)^{nm+m(m-1)/2}(\delta_i,\delta_j)\,\,\,\,\mbox{for
$i\ne j$.}
\end{equation}
Formula (\ref{eq9}) is simplified for even $m$,
then $(\widetilde\delta_i,\widetilde\delta_j) =
(-1)^{m/2} (\delta_i, \delta_j)$.
In what follows we identify the Milnor lattices of the critical
points $f$ and $f\oplus Q$ and thus elements of a distinguished
basis of $L_f$ and of the corresponding distinguished basis of
$L_{f\oplus Q}$ (using the same notations for them).

Formula (\ref{eq9}) shows that there are only four
different intersection forms of
stabiliations. The two of the four forms, corresponding to
stabilizations with an odd number of variables, are symmetric and the two,
corresponding to stabilizations with an even number of
variables, are skew-symmetric. The two forms of the same
type (symmetric or skew-symmetric) differ only by the common
sign.

Let the number of variables, $n+m$, satisfy
$ n+m \equiv 1\ {\rm{mod}}\,4$. Then the corresponding
intersection form will be called the quadratic form of
the germ $f$ (or the symmetric bilinear form).
We denote it by $(\cdot,\cdot)_q$.
In this case the square of each vanishing cycle
equals 2.

Given a distinguished basis $\{\delta_i\}$, the quadratic form
can be described by a weighted graph, the Dynkin diagram
of $f$. The vertices of the graph are elements of the distinguished
basis. Any two vertices $\delta_i$ and $\delta_j$
are connected by an edge if and only if
$(\delta_i, \delta_j)_q \ne 0$. If $(\delta_i, \delta_j)_q
\ne 0$, then
the vertices are connected by an edge
of multiplicity $-(\delta_i, \delta_j)_q$.

The monodromy groups of stabilizations with an
odd number of variables are naturally isomorphic.
The isomorphism identifies the corresponding
classical monodromy operators.
The same is true for the monodromy groups of stabilizations
with an even number of variables.

The monodromy group of a stabiliation with an odd number of variables
is a group generated by reflections. The group will be called
the reflection group of the germ $f$. In this case the operator
of classical monodromy will be called the Coxeter element of $f$.

Gabrielov's theorem implies the following description of the quadratic
form of the direct sum $f\oplus g$ of the germs $f$ and $g$.
It is naturally isomorphic to the quadratic form on the tensor
product $L_f\otimes L_g$ of the Milnor lattices $L_f$ and $L_g$
defined by the formulae ($\Delta_{ij}=\delta_i\otimes\delta_j^\prime$):
$$
\begin{array}{ll}
(\Delta_{i\,j_1},\Delta_{i\,j_2})_q=
(\delta_{j_1}^\prime,\delta_{j_2}^\prime)_q&{\mbox{for $j_1\ne j_2$,}}
\\
(\Delta_{i_1j\,},\Delta_{i_2j\,})_q=
(\delta_{i_1},\delta_{i_2})_q&{\mbox{for $i_1\ne i_2$,}}
\\
(\Delta_{i_1j_1},\Delta_{i_2j_2})_q=0&{\mbox{for $(i_2-i_1)(j_2-j_1)<0$,}}
\\
(\Delta_{i_1j_1},\Delta_{i_2j_2})_q=
(\delta_{i_1},\delta_{i_2})_q(\delta_{j_1}^\prime,\delta_{j_2}^\prime)_q
&{\mbox{for $(i_2-i_1)(j_2-j_1)>0$,}}
\\
(\Delta_{ij},\Delta_{ij})_q=2.
&{}
\end{array}
$$

These formulae imply a description of the corresponding Dynkin
diagram of the direct sum of two germs.

Let $f_t(x)$ be a continuous family of germs
of functions having an isolated critical point.
Assume that the Milnor number of the critical point
does not depend on $t$.
Then the Milnor lattices, the quadratic forms, the reflection groups,
and the Coxeter elements of all
the germs $f_t$ are naturally isomorphic.
Moreover, they have the same Dynkin diagrams.

For instance, consider the family
of germs at the origin of quasi-homogene\-ous polynomials of a given degree,
given weights of variables, and having an isolated critical
point. Then the Milnor lattices, the quadratic forms,
the reflection groups, and the Coxeter elements of all
the germs are naturally isomorphic.

\section{Multiplication in the Verlinde algebra, to\-p\-ology of
the short fusion potential, and the level-rank duality}\label{sec4}

\begin{theorem}\label{theo3}
For any $N >1$ and $k\geq 0$, consider the Verlinde algebra
of
$su(N)_k$ and the critical point
of the short fusion potential $V^0_{N,k}(x_1,...,x_{N-1})$.
Then there is an isomorphism of the lattice
of the Verlinde algebra and the Milnor lattice
of the critical point,
$$
\psi \,:\, L(su(N)_k) \, \to \, L_{V^0_{N,k}},
$$
sending the bilinear form $B_{n,k}$ to
the quadratic form on the Milnor lattice
and sending the basis $\alpha_\lambda,\, \lambda \in
P_{N,k}$, with the lexicographical ordering to
a distinguished basis of the Milnor lattice.
\end{theorem}

\begin{corollary}\label{cor1}
The isomorphism $\psi$ sends the reflection group
$\Gamma_{N,k}$ and the Coxeter element $M_{N,k}$
to the reflection group of the critical point and
its Coxeter element, respectively.
\end{corollary}

\begin{theorem}\label{theo4}
Let $N\geq k$. Then there is a continuous family
of germs of holomorphic functions at an isolated critical point
with a constant Milnor number,
$f_s\, :\,(\C^N,0) \, \to\, (\C,0),\, s \in [0,1]$,
such that the germ $f_0$ is the germ
of the short fusion potential $V^0_{N+1,k}(x_1, ..., x_{N})$ at the
origin and the germ $f_1$ is stable equivalent to the germ
at the origin of another short fusion potential $V^0_{k+1,N}(x_1,
..., x_{k})$.
\end{theorem}

\begin{corollary}\label{cor2}, {\bf Level-Rank Duality.}
There exists an isomorphism
$\psi:
L(su(N+1)_k)  \to  L(su(k+1)_N)$
of the lattices
sending the bilinear form $B_{N+1,k}$, the reflection
group $\Gamma_{N+1,k}$, and the Coxeter element $M_{N+1,k}$
to the bilinear form $B_{k+1,N}$, the reflection
group $\Gamma_{k+1,N}$, and the Coxeter element $M_{k+1,N}$,
respectively.
\end{corollary}

Theorem 3 allows us to apply results on Milnor
lattices of critical points to the lattices of Verlinde
algebras, in particular, see in {\cite{E}} a description
of the reflection groups of critical points and in
{\cite{S}} a formula for the signature of the intersection
form of a quasi-homogeneous critical point.

Theorem 3 and Corollaries 1, 2 were conjectured by J.-B. Zuber
in {\cite{Z}}.
The level-rank duality conjecture was motivated in {\cite{Z}}
by analogies with $N=2$ superconformal theories.
Theorem 3 and Corollary 2 were proved for $N=3$ by
N. Warner {\cite{Wa}}. Warner's proof helped us to invent a proof
for general $N$. Theorem 4
for $N=k+1$ could be recognized in {\cite{Z}}.

\section{Proofs}\label{sec5}

\subsection{Lemmas}

The following two lemmas will be used in the proof of Theorem 3.

It is well-known {\cite{AGV}} that the critical point of the germ
$y^h$ of type $A_{h-1}$ has a distinguished basis
$\delta_i$, $i=1, 2, \ldots, h-1$,
with the intersection numbers $(\delta_i, \delta_i)=2$, and
$(\delta_i, \delta_j)=0$ for
$\vert j-i\vert>1$, $(\delta_i, \delta_{i+1})=1$.
Any distinguished basis with this property will be called
special.

\begin{lemma}\label{lem1}
There exists a continuous
family $g_s(y)$, $s\ge0$, of polynomials of degree $h$
with the coefficients of $y^h$ and $y^{h-1}$ equal to 1 and 0,
respectively, and such that:
\newline
1) $g_s(y)$ is a deformation of $y^h$, i.e. $g_0(y)=y^h$;
\newline
2) all the critical
points $y^{(j)}$, $j=1, 2, \ldots, h-1$, of $g_s$ are non-degenerate
and the critical values $g_s(y^{(j)})$ are
pair-wise distinct;
\newline
3) the critical value $g_s(y^{(j)})$ is equal to
$({1\over 3N})^j s$, where $j=1,...,h-1;$
\newline
4) fix a non-critical value $z^{(0)}$ in the upper
half-plane, consider the system of intervals
$u_j(t)=tz^{(0)}+(1-t)g_s(y^{(j)})$ connecting
the non-critical value and the critical values,
then the distinguished basis corresponding to
this system of paths is special.
\end{lemma}

{\bf Proof of Lemma 1.}
Let $\C_a^{h-1}$ be the space of polynomials of the
form $y^h+a_{h-2}y^{h-2}+\ldots+a_1y+a_0$, $a=(a_0, a_1,\ldots a_{h-2})$.
Let $\Sigma\subset\C_a^{h-1}$ be
the subset of polynomials with multiple
critical values, i.e.
 the polynomials having a degenerate critical point or
a pair of critical points with equal critical values.

Let $\C^{h-1}_z$ be the space of unordered sets of $h-1$ complex numbers
$(z_1$, \dots, $z_{h-1})$.
Let $\Delta\subset \C^{h-1}_z$ be the subspace of unordered sets
$(z_1,$ \dots, $ z_{h-1})$ with $z_i=z_j$ for some $i$ and $j$.

Let $p:\C_a^{h-1}\to \C_z^{h-1}$ be the map sending a polynomial
to the set of its critical values. $p$ maps the complement
of $\Sigma$ to the complement of $\Delta$.
By the O.Lyashko and E.Looijenga theorem ({\cite{A}}, {\cite{L}})
the map $p$ is proper (i.e. the preimage of a compact subspace is
compact), the preimage of 0 consists of one point corresponding to
 the polynomial
$y^h$, and the restriction of $p$ to $\C_a^{h-1}\setminus\Sigma$
is a covering map over $\C^{h-1}_z\setminus\Delta$.

Let $f(y)\in\C_a^{h-1}$ be a deformation of $y^h$ with
nondegenerate critical points and pair-wise
distinct critical values $z_1$, \dots $z_{h-1}$.
 Fix a non-critical value $z_0$ of $f$ lying
in the upper half-plane. Fix a
system of paths $u_i(t)$, $i=1, \ldots, h-1$,
$u_i(0)=z_i$, $u_i(1)=z_0$, connecting
the non-critical value with the critical values and defining
a special distinguished basis of vanishing cycles $\delta_1$, \dots,
$\delta_{h-1}$.
 We have
$p(f)=(z_1, \ldots, z_{h-1})$.

For $\tau\in [0,1)$ let $z_i(\tau)=u_i(\tau)$,
and $u_{\tau, i}(t)=u_i((1-\tau)t+\tau)$. The family
of unordered sets
$(z_1(\tau), \ldots, z_{h-1}(\tau))$
of points of the space $\C_z^{h-1}$ is a homotopy of
the point $(z_1, \ldots, z_{h-1})$.

For $\tau_0$ close to 1,
the paths $u_{\tau_0, i}$ are close to straight lines.
The system of points $z_1(\tau_0)$, \dots, $z_{h-1}(\tau_0)$
in $\C$ and paths $u_{\tau_0, i}$
can be easily deformed into the system described in Lemma 1.
By the theorem of O.Lyashko and E.Looijenga
every homotopy of $h-1$ distinct points in $\C$ gives rise
to a unique homotopy of the corresponding polynomial.
So deforming the initial system of paths $\{u_i\}$
into the system of paths described in section 4 of Lemma 1
we construct a deformation of the initial polynomial $f(y)$
to a polynomial which we denote by $g_s(y)$. This polynomial
considered with the system of paths described in section 4 of Lemma 1
satisfies the conditions of Lemma 1. The fact
that $g_s(y)$ tends to $y^h$ for $s\to 0$ follows from the
fact that the map $p$ is proper. Lemma 1 is proved.
 $\Box$
\smallskip

Let $\C^{n}$ be the complex linear space with the standard $S_n$-action,
$C\subset\C^{n}$ the union of non-regular orbits
of the action, i.e. the union of its mirrors.
Let $f$ be an $S_n$-invariant holomorphic function defined
in a neighbourhood of a point $y\in C$.
Let $B_\delta(y)$ be the ball of radius $\delta$ in $\C^n$ with the
center at $y$, $D_\eps$ the disk of radius $\eps$ in $\C$ with the
center at $f(y)$.
Suppose that the point $y$ is
not a critical point of the function $f$.
Then there exist positive $\delta$ and $\eps$ such that
the restriction of $f$ to $B_\delta(y)\cap f^{-1}(D_\eps)$
is a locally trivial bundle over $D_\eps$.

\begin{lemma}\label{lem2}
There exist positive $\delta$ and $\eps$ such that $f$ defines
a locally trivial bundle of pairs,
$f:(B_\delta(y)\cap f^{-1}(D_\eps), C\cap B_\delta(y)\cap f^{-1}(D_\eps))\to
D_\eps$.
\end{lemma}

{\bf Proof of Lemma 2.}
$C$ is the union of hyperplanes in $\C^n$,
the mirrors $\{y_i=y_j\}$ of the $S_n$-action.
In order to prove that $f$ defines a locally trivial bundle
of pairs it suffices to show that
the restriction of $f$ to any intersection $M$ of some mirrors
does not have a critical point at $y$. Suppose that $y$ is a critical
point of $f\vert_ M$, i.e. $d\,f\vert_ M=0$ at $y$.
Let $G$ be the subgroup
of $S_n$ generated by
the reflections at the
 mirrors containing $M$. Let $M^\perp$ be the orthogonal
complement to $M$ at $y$.
 The action of $G$ on $M^\perp$
is a group generated by reflections. The intersection
of the corresponding mirrors in $M^\perp$ is trivial, coincides with
$y$.

Let a linear function $\ell$ on $M^\perp$ be invariant with respect
to the action of $G$. Then it is invariant with respect to each reflection
in $G$ and thus, its gradient lies in the corresponding mirror. This
implies that $\ell=0$.

The restriction $f\vert_{ M^\perp}$ is an
$G$-invariant function on $M^\perp$ and thus, its differential
$d\,f\vert_{M^\perp}$ at $y$ equals zero. By assumptions we
have $d\,f\vert_{M}=0$
and $d\,f\vert_{M^\perp}=0$ at $y$. Hence, $d\,f\vert_{ y}=0$ and $y$
is a critical point of $f$. Lemma 2 is proved. $\Box$
\smallskip

\subsection{Proof of Theorem 3}

The symmetric group $S_{N-1}$ acts on the space $\C_y^{N-1}$
permuting coordinates. Let
$$
\pi : \C_y^{N-1} \to \C_x^{N-1}, \qquad (y_1,...,y_{N-1})
\mapsto (x_1,...,x_{N-1}),
$$
be the quotient map, where $x_n$ are the elementary symmetric
functions defined in (\ref{eq4}).

Let $h=N+k$. Consider the lifting of the short fusion potential
to the preimage of $\pi$,
$ f(y)=V^0_{N, k}(\pi(y)) = (y_1^h + y_2^h +
\ldots + y_{N-1}^h)/h$.
The Milnor fiber of the germ $f$ is invariant with
respect to the $S_{N-1}$-action and thus, the group $S_{N-1}$
acts on the Milnor lattice of $f$. The map $\pi$ maps the Milnor fiber
of $f$ onto the Milnor fiber of $V^0_{N, k}$.

The Milnor lattice of
the germ $f(y)$ is the tensor product of $N-1$ copies
of the Milnor lattice of $y^h$ (section 3).
A permutation $\sigma$ from the group $S_{N-1}$ acts on the tensor product
permuting the factors and multiplying the result by $(-1)^{\eps(\sigma)}$,
where $\eps(\sigma)$ is the parity of $\sigma$. According to Gabrielov's
theorem {\cite{Gab}} the germ $f$ has a distinguished basis
consisting of the tensor products of the basis
vanishing cycles of the germ $y^h$, $\Delta_{\bar\imath}=
\delta_{i_1}\otimes\ldots\otimes\delta_{i_{N-1}}$,
where $\bar\imath=(i_1, \ldots, i_{N-1})\in \{1, 2, \ldots, h-1\}^{N-1}$,
ordered lexicographically. Namely,
$\bar\imath=(i_1, \ldots, i_{N-1})<\bar\jmath=(j_1, \ldots, j_{N-1})$,
if there exists $n$ such that $i_m=j_m$ for $m>n$ and $i_n<j_n$.
To describe the quadratic form of $f$ on the Milnor lattice
 we describe the corresponding bilinear symmetric
form by the rule: for
${\bar\imath}< {\bar\jmath}$, we have
$(\Delta_{\bar\imath},\Delta_{\bar\jmath})_q=1$
if $j_n$ equals either $i_n$ or $i_n+1$ for all
$n=1, 2, \ldots, N-1$, and
$(\Delta_{\bar\imath},\Delta_{\bar\jmath})_q=0$
otherwise.

Denote by $Q_{N,k}$ the set $\{\bar\imath=(i_1,\ldots,i_{N-1}):
1\le i_{N-1}<\ldots<i_1\le h-1\}$ and by $\bar Q_{N,k}$
the set $\{\bar\imath: 1\le i_{N-1}\le\ldots\le i_1\le h-1\}$.

\begin{lemma}\label{lem3}
The cycles $\widetilde\Delta_{\bar\imath}=\pi_\star(\Delta_{\bar\imath})$
with ${\bar\imath}\in Q_{N, k}$ form a distinguished basis of the
Milnor lattice of the short fusion potential $V^0_{N, k}(y)$. Moreover
the cycles $\Delta_{\bar\imath}$ with ${\bar\imath}\in Q_{N, k}$
can be realized geometrically so that
$\Delta_{\bar\imath}\cap C=\emptyset$.
\end{lemma}

We finish the proof of Theorem 3 and then prove Lemma 3.

The intersection number of the vanishing cycles
$\widetilde\Delta_{\bar\imath}$,
$\widetilde\Delta_{\bar\jmath}$ of the short fusion potential
for
 ${\bar\imath}, {\bar\jmath}\in Q_{N, k}$ can be obtained
in the following way. By Lemma 3 the cycles
$\Delta_{\bar\imath}$ and
$\Delta_{\bar\jmath}$ can be realized geometrically so that
$\Delta_{\bar\imath}\cap C=\emptyset$ and
$\Delta_{\bar\jmath}\cap C=\emptyset$.
The intersection points of the geometric
cycles $\widetilde\Delta_{\bar\imath}$ and $\widetilde\Delta_{\bar\jmath}$
do not lie in $\pi (C)$ and
are the images under $\pi$ of the intersection points of the
cycle $\Delta_{\bar\imath}$ with all the cycles of the form
$\sigma \Delta_{\bar\jmath}$ for $\sigma\in S_{N-1}$.
Thus,
$(\widetilde\Delta_{\bar\imath}, \widetilde\Delta_{\bar\jmath})=
\sum_{\sigma \in S_{N-1}}(\Delta_{\bar\imath},
\sigma\Delta_{\bar\jmath})
$. It is not difficult to see that for ${\bar\imath}$ and ${\bar\jmath}$
in $Q_{N, k}$ and $\sigma\ne 1$ we have
$(\Delta_{\bar\imath}, \sigma\Delta_{\bar\jmath})=0$.
Hence $(\widetilde\Delta_{\bar\imath}, \widetilde\Delta_{\bar\jmath})=
(\Delta_{\bar\imath}, \Delta_{\bar\jmath})$ and thus, the Dynkin diagram
of the short fusion potential $V_{N, k}(x)$ in the distinguished
basis $\{\widetilde\Delta_{\bar\imath}\}$ is a part of the described
Dynkin diagram of the function $f$, the part corresponding to the
vanishing cycles $\Delta_{\bar\imath}$ , ${\bar\imath} \in Q_{N, k}$.
In particular this implies that
$(\widetilde\Delta_{\bar\imath}, \widetilde\Delta_{\bar\jmath})_q=
(\Delta_{\bar\imath}, \Delta_{\bar\jmath})_q$ (${\bar\imath},
\bar\jmath \in Q_{N, k}$).

For $\lambda \in P_{N,k}$, set
$$
\bar\imath(\lambda)=
(1+\lambda_1+\lambda_2+\ldots+\lambda_{N-1},
1+\lambda_2+\ldots+\lambda_{N-1}, \ldots,
1+\lambda_{N-1}) \, \in Q_{N,k}.
$$
It is not difficult to see that the correspondence
$\alpha_{\lambda}\mapsto \widetilde\Delta_{{\bar\imath}(\lambda)}$
between the basis elements
$\alpha_\lambda$, $\lambda \in P_{N,k}$, of the lattice of the Verlinde
algebra of $su(N)_k$
and the basis vanishing cycles $\{\widetilde\Delta_{\bar\imath}\}$,
$\bar\imath\in Q_{N,k}$, defines the isomorphism of Theorem 3. $\Box$

\smallskip
{\bf Proof of Lemma 3.}
Fix a small $s \in (0,1]$ and let
$g(y)=g_s(y)$ be the deformation of the function
$y^h$ described in Lemma 1. Set
$\widetilde f(y)={1\over h}\sum_{i=1}^{N-1}g(y_i)$.
The function $\widetilde f$ is an $S_{N-1}$-invariant
deformation of $f(y)$
with only non-degenerate critical points. Moreover,
$\widetilde f$ has different critical values at different
orbits of critical points. The critical points
of $\widetilde f$ are the points
$p_{\bar\imath}=(y^{(i_1)}, y^{(i_2)}, \ldots,
y^{(i_{N-1})})$ for
$\bar\imath=(i_1, \ldots, i_{N-1})\in \{1, 2, \ldots, h-1\}^{N-1}$.
All the critical values
$z_{\bar\imath}={1\over h}\sum_{j=1}^{N-1}g(y^{(i_j)})$
are real and if $\bar\imath$ and $\bar\jmath$ are in $\bar Q_{N, k}$
and ${\bar\imath}<{\bar\jmath}$, then $z_{\bar\imath}>z_{\bar\jmath}$.

Let $\widetilde V(x)$ be the deformation of the short fusion potential
$V^0_{N, k}$ such that $\widetilde f(y) = \widetilde V(\pi (y))$.
The critical points of $\widetilde V$
correspond to the critical points $p_{\bar\imath}$ of $\widetilde f$
with ${\bar\imath}\in Q_{N, k}$.
Take the number ${N-1\over h}z^{(0)}$
as a non-critical value of $\widetilde f$
(and thus, of $\widetilde V$),
see Lemma 1. It follows from the
proof of the Gabrielov theorem \cite{Gab} that the vanishing cycle
$\Delta_{\bar\imath}$ with ${\bar\imath}\in \bar Q_{N, k}$ vanishes
along the path $v_{\bar\imath}(t)$, $0\le t\le N-1$, defined by the formula:
$v_{\bar\imath}(t)={n\over h}z^{(0)}+{1\over h}\left(\sum_{j=1}^{N-n-2}
g(y^{(i_j)})+u_{i_{N-n-1}}(t-n)\right)$
for $n\le t\le n+1$. The path $v_{\bar\imath}(t)$ is composed from
shifts of the corresponding intervals $u_{i_n}$.

The paths $v_{\bar\imath}$, ${\bar\imath}\in\bar Q_{N, k}$, can be
deformed inside
the upper half-plane
so that they will not intersect each other except
at their end point
${N-1\over h}z^{(0)}$. The paths $v_{\bar\imath}$ with
${\bar \imath }\in Q_{N, k}$ as well as their small deformation
 do not go through the critical values of
$\widetilde f$ at the critical points lying
on the union of mirrors, i.e. through
$z_{\bar\jmath}$ with ${\bar\jmath}\in\bar Q_{N, k}\setminus Q_{N, k}$.

For any
$ \imath
\in
Q_{N,k}$ the union of mirrors
$C$ forms a subbundle in the bundle of local level manifolds
of $\widetilde f$ over the path $v_{\bar\imath}$, see Lemma 2. Therefore,
the spheres defining the vanishing cycle $\Delta_{\bar\imath}$ (and
thus, the cycle $\widetilde
\Delta_{\bar\imath}$ itself) can be chosen in such a way
that they do not intersect $C$. It implies that
the cycle $\widetilde\Delta_{\bar\imath}=\pi(\Delta_{\bar\imath})$
vanishes along the path $v_{\bar\imath}$. The system of paths
$\{v_{\bar\imath}(t):\, {\bar\imath}\in Q_{N, k}\}$ satisfies
the conditions for a system of paths to define a distinguished
basis of vanishing cycles. This implies the statement of Lemma
3. $\Box$

\subsection{Proof of Theorem 4}

Consider a function
$$
F(x_1, \ldots, x_N)=V^0_{k+1, N}(x_1, \ldots, x_k)+
\sum_{i=k+1}^Nx_ix_{N+k-i+1},
$$
which is a stabilization of the short fusion potential
$V^0_{k+1,N}(x_1, ..., x_{k})$.
Both functions: the function
$F(x_1, ..., x_N)$ and the short fusion potential
$V^0_{N+1,k}(x_1, ..., x_{N})$ are quasi-homogeneous polynomials
of degree $N+k+1$ with the weight of the variable $x_n$ equal to $n$,
$n=1,..., N$. Both functions have an isolated critical point at the origin.
This proves Theorem 4. $\Box$

\enddocument